\documentclass[10pt]{article}
\usepackage[top=0.5in,left=1.5in,footskip=0.5in]{geometry}

\usepackage[utf8]{inputenc}

\usepackage{cite}

\usepackage{enumerate}

\usepackage{nameref,hyperref}

\usepackage{microtype}
\DisableLigatures[f]{encoding = *, family = * }

\usepackage{changepage}

\usepackage[aboveskip=1pt,labelfont=bf,labelsep=period,singlelinecheck=off]{caption}

\makeatletter
\renewcommand{\@biblabel}[1]{\quad#1.}
\makeatother

\usepackage{color}

\definecolor{Gray}{gray}{.25}

\usepackage{graphicx}

\usepackage{sidecap}

\usepackage{amsmath}
\usepackage{bm}
\usepackage{amsfonts}
\usepackage{mathrsfs}
\usepackage{stmaryrd} 

\newtheorem{theorem}{Theorem}
  \newcommand{\qedd}{\hfill \Box}

\newtheorem{definition}{Definition}
\newtheorem{assumption}{Assumption}
\newtheorem{lma}{Lemma}

\usepackage{wrapfig}
\usepackage[pscoord]{eso-pic}
\usepackage[fulladjust]{marginnote}
\reversemarginpar

\begin{document}
\vspace*{0.35in}

\begin{flushleft}
{\Large
\textbf\newline{Positive Trigonometric Polynomials on the Stability of Spatially Interconnected Systems}
}
\newline
\\
Xiaokai Zhai\textsuperscript{1,*}
\\
\bigskip
\bf{1} School of Mathematical Sciences, Suzhou University of Science and Technology, Suzhou 215009, Jiangsu Province, PR China
\\
\bigskip
* xkzhai@usts.edu.cn

\end{flushleft}

\justifying\let\raggedright\justifying
\raggedright

\section*{Abstract}
This paper is devoted to the stability analysis of spatially interconnected systems (SISs) via the sum-of-squares (SOS) decomposition of positive trigonometric polynomials. For each spatial direction of SISs, three types of interconnected structures are considered. Inspired by the idea of rational parameterization and robust stabilizability function, necessary and sufficient conditions are derived for establishing the stability of SISs with two different combined topologies respectively. For these results, the primary issue concerns the global or local positivity of trigonometric polynomials. SOS decomposition and generalized trace parameterization of positive trigonometric polynomials are utilized so that the addressed problems can be quantified by two semidefinite programs (SDPs). The proposed methods are applicable to all possible interconnected structures due to the assumption of spatial reversibility. Numerical examples are given to illustrate the efficiency of the derived theoretical results.  


\section{Introduction}
Spatially interconnected systems (SISs), also referred to as spatially distributed systems, are generally regarded as large-scale interconnected systems consisting of multiple spatially distributed similar subunits that only interact with their neighbors. Many practical applications, such as heat equation \cite{Bamieh2002}, vehicle platoons \cite{Knorn2013}, and flexible structures \cite{Liu2016},  can be captured by such interconnected systems. There are also many other excellent state-space representations in the literature can be used to describe the dynamic of these systems such as Roesser model \cite{Roesser1975,Ahn2015}, Fornasini-Marchesini model \cite{Fornasini1976,Zhang2017}, and multidimensional (MD) model \cite{Xu2012,Wang2017}, etc.  The SISs model is originally proposed in \cite{DAndrea2003}, where the physical relevance is sufficiently taken into account. As a result, not only causal but also noncausal spatial dynamics appear in SISs, leading to a distinct framework on system description. Considerable efforts have been made under this framework \cite{AlTaie2016,Zhou2008,Kim2013,Heijmans2017,Xu2018}. 

One of the primary problems in SISs is stability analysis, for which a general accepted method consists of recasting SISs as an infinite-dimensional system \cite{Curtain1995,Bamieh2002}, and employing the Lyapunov theory of distributed parameter systems to obtain linear operator inequalities (LOIs) analysis conditions \cite{DAndrea2003,Dullerud2004,Recht2004,Fridman2009}. However, it must be mentioned that the existing LOIs results are generally sufficient, but not necessary for stability analysis (and accordingly, for controller synthesis).  In \cite{Zhou2008}, the idea of parameter-dependent linear matrix inequalities is employed to derive necessary and sufficient analysis conditions for SISs, whereas it remains challenging to find an efficient method for determining the degree of the related matrix polynomials. Recently, the sum-of-squares (SOS) decomposition of positive polynomials has been intensively investigated, and widely utilized by the authors from \cite{Chesi2010,Chesi2014,Chesi2016,Chesi2019} to obtain nonconservative analysis conditions for 2D mixed continuous-discrete-time systems. These encouraging results motivate us to develop their counterparts in SISs, which is not a trivial work. As mentioned above, the physical relevance of SISs results in an essential difference for stability analysis. Mathematically, the ordinary polynomial is replaced by polynomials in complex parameters over unit circle. Moreover, the arbitrary number of spatial directions converts univariate polynomials to multivariate case, imposing some limitations on the SOS decomposition.

In this work, the well-established theory of trigonometric polynomials is utilized to address these problems via SOS decomposition of trigonometric counterparts \cite{Dumitrescu2006,Dumitrescu2007,Dumitrescu2017}. The contribution of this note is threefold. Firstly, a necessary and sufficient stability condition is derived for SISs of full infinite interconnections, including a stability test of a constant matrix, and a global positivity of a trigonometric polynomial over unit $L$-circle. Secondly, the stability of SISs with mixed infinite and periodic interconnections is established via Routh-Hurwitz criterion, which contains a check of empty set and local positivity of a series of trigonometric polynomials over domains.  Finally, with the aid of the generalized trace parameterization of trigonometric polynomials, two semidefinite programs (SDPs) are proposed to quantify the derived theoretical results.

The rest of this paper is organized as follows. Preliminaries and the problem formulation are introduced in Section \ref{sec2}. Section \ref{sec3} derives main results of this note, and numerical examples are presented in Section \ref{sec4} to illustrate the proposed methodologies.  Conclusions are included in Section \ref{sec5}.

\emph{Notation.} The notation used throughout is reasonably standard. $\mathbb{Z}$ stands for the set of integers, and the superscript `$L$' denotes the Cartesian product of $L$ identical sets.  The notation $\mathbb{T}$ is used to indicate the unit circle, i.e, $\mathbb{T}=\left\{z\in \mathbb{C}:~|z|=1\right\}$. The set of non-negative real numbers and complex numbers are denoted by $\mathbb{R}^+$ and $\mathbb{C}$, respectively.  $\mathbb{R}^\bullet$ denotes the real-valued vector whose size is not relevant to the discussion, where $\mathbb{R}$ is the set of real numbers. The set of  $n\times m$ complex matrices and real matrices is represented by $\mathbb{C}^{n\times m}$ and $\mathbb{R}^{n\times m}$, respectively. $\lfloor \cdot \rfloor$ and $\lceil \cdot \rceil$ denote the floor and ceiling operators, respectively. Let $M\in \mathbb{C}^{n\times m}$ be a given complex matrix, $M^T$ and $M^*$ are used to represent its transpose and complex conjugate transpose, respectively. For matrices $M$ and $N$, $M\otimes N$ indicates the Kronecker product of them.  The trace of matrix $M$ is denoted by $tr(M)$. For scalars $a$, $b$, $c$, $d$,  the notation $\left \llbracket \cdot \right \rrbracket$ denotes 
\begin{equation}
\left \llbracket 
\begin{array}{cc}
a & b\\
c & d
\end{array}
 \right \rrbracket = \frac{bc - ad}{c}.
\end{equation}
Finally, the  vector $[x_1^*~~x_2^*]^*$ will be described as the notation $(x_1;x_2)$ for simplicity.

\section{Preliminaries} \label{sec2}
\subsection{System model and Problem formulation}

In this paper, the signals considered are vector valued functions indexed by $L+1$ independent variables, i.e, $x=x(t,k_1,...,k_L)$, in which $t\in \mathbb{R}^{+}$ denotes the temporal variable, and $k_i \in \mathbb{D}_i$ denotes the $i$-th spatial variable, where $\mathbb{D}_i$ stands for any one of the following three sets: $\mathbb{Z}$ for infinite spatial extent in $i$-th dimension, $\mathbb{Z}_{N_i}$ for periodicity of period $N_i$, and $\left\{1,...,N_i \right\}$ for finite extent interconnection. 

According to the structure of signals, the following definitions are first recalled. The $L$-tuple $(k_1,...,k_L)$ is abbreviated as $\mathbf{k}$ for notational simplicity. 
\begin{definition} \cite{DAndrea2003}
	The space $\ell_2$ is the set of functions $x$ mapping $\mathbb{D}_1\times \cdots \times \mathbb{D}_L$ to $\mathbb{R}^{\bullet}$ for which the following inequality is satisfied:\begin{equation}
	\sum_{k_1\in \mathbb{D}_1}\cdots \sum_{k_L\in \mathbb{D}_L}x^*(\mathbf{k})x(\mathbf{k})<\infty.
	\end{equation}
	The inner product on $\ell_2$ is defined as 	\begin{equation}
	{\left\langle {x,y} \right\rangle _{{\ell _2}}} := \sum\limits_{{k_1} \in \mathbb{D}_1}  \cdots  \sum\limits_{{k_L} \in \mathbb{D}_L} {{x^*}(\mathbf{k})y(\mathbf{k})},
	\end{equation}
	with corresponding norm $\|x\|_{\ell_2}:=\sqrt{\langle x,x \rangle_{\ell_2}}$.
\end{definition}

For a fixed temporal variable $t$ and spatial variables $\mathbf{k}$, $x(t,\mathbf{k})$ is a real-valued vector, and $x(t)$ denotes a signal in $\ell_2$.

The following continuous-time SISs is considered in this paper:
\begin{equation}\label{Sigma1}
\begin{aligned}
(\Sigma):~\begin{bmatrix}
\frac{\partial x(t,\mathbf{k})}{\partial t}\\ w(t,\mathbf{k}) 
\end{bmatrix}&=\begin{bmatrix}
A_{TT} & A_{TS}\\A_{ST} & A_{SS}
\end{bmatrix}\begin{bmatrix}
x(t,\mathbf{k}) \\ v(t,\mathbf{k})
\end{bmatrix},\\
x(0,\mathbf{k}) & = x_{0}(\mathbf{k}) ,
\end{aligned} 
\end{equation}
where $x(t,\mathbf{k})\in \mathbb{R}^{n_0}$ denotes the state vector, $w(t,\mathbf{k})$, $v(t,\mathbf{k})\in \mathbb{R}^{n}$ are interconnection variables between subsystems with the forms of
\begin{equation*}
 \begin{aligned}
v(t,\mathbf{k}) &= \left(v_{1}(t,\mathbf{k});v_{-1}(t,\mathbf{k});\cdots;v_{L}(t,\mathbf{k});v_{-L}(t,\mathbf{k})\right),\\
w(t,\mathbf{k}) &= \left(w_{1}(t,\mathbf{k});w_{-1}(t,\mathbf{k});\cdots;w_{L}(t,\mathbf{k});w_{-L}(t,\mathbf{k})\right),
\end{aligned}
\end{equation*}
in which, $v_{i}(t,\mathbf{k})$, $w_{i}(t,\mathbf{k}) \in \mathbb{R}^{n_i}$, $v_{-i}(t,\mathbf{k})$, $w_{-i}(t,\mathbf{k})\in \mathbb{R}^{n_{-i}}$ ($i=1,...,L$), and $\sum_{i=1}^{L}(n_i+n_{-i})=n$. 
As indicated by the different definitions of $\mathbb{D}_i$,  three types of spatial interconnections are considered for $i$-th spatial direction.
\begin{enumerate}[(1)]
	\item Infinite Interconnection ($\mathbb{D}_i=\mathbb{Z}$):
	\begin{equation}
	\begin{cases}
		v_{i}(t,\mathbf{k}|_{k_i=l+1})=w_i(t,\mathbf{k}|_{k_i=l}),~\forall  l \in \mathbb{Z},\\
	v_{-i}(t,\mathbf{k}|_{k_i=l-1}) = w_{-i}(t,\mathbf{k}|_{k_i=l}),~\forall  l \in \mathbb{Z}.
	\end{cases}
	\end{equation}
	\item Periodic Interconnection ($\mathbb{D}_i=\mathbb{Z}_{N_i}+1$):
		\begin{equation}
		\begin{cases}
		v_{i}(t,\mathbf{k}|_{k_i=l+1}) = 	w_i(t,\mathbf{k}|_{k_i=l}),~1\leq l \leq N_i - 1,\\
		 v_{-i}(t, \mathbf{k}|_{k_i=l-1})=w_{-i}(t, \mathbf{k}|_{k_i=l}),~2 \leq l \leq N_i,\\
		 	v_{i}(t,\mathbf{k}|_{k_i=1}) = 	w_i(t,\mathbf{k}|_{k_i=N_i}),\\
		 	 v_{-i}(t, \mathbf{k}|_{k_i=N_i})=w_{-i}(t, \mathbf{k}|_{k_i=1}).
		\end{cases}
	\end{equation}
		\item Spatially $M$-reversible finite extent system ($\mathbb{D}_i = \left\{ 1,...,N_i\right\}$):
	\begin{equation}
	\begin{cases}
	v_{i}(t,\mathbf{k}|_{k_i=l+1})=w_i(t,\mathbf{k}|_{k_i=l}),~1\leq l \leq N_i -1,\\
	v_{-i}(t, \mathbf{k}|_{k_i=l-1})=w_{-i}(t, \mathbf{k}|_{k_i=l}),~2\leq l \leq N_i,\\
	v_i(t,\mathbf{k}|_{k_i=1}) = M_iw_{-i}(t,\mathbf{k}|_{k_i=1}), \\
		v_{-i}(t,\mathbf{k}|_{k_i=N_i}) = M_i^{-1}w_i(t,\mathbf{k}|_{k_i=N_i}). \\
	\end{cases}
	\end{equation}
\end{enumerate}
In which, $$\mathbf{k}|_{k_i=l}=(k_1,...,k_{i-1},l,k_{i+1},...,k_L),$$
$M_i$ is a nonsingular boundary conditions matrix for finite spatial extent in dimension $i$ (called the boundary conditions matrix), and the finite extent system is restricted to be spatially $M$-reversible (see for details \cite{Langbort2005}). 

\begin{figure}[!t]
	\centering
		\includegraphics[width=0.4in]{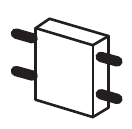}
		\hfil
	\includegraphics[width=1.5in]{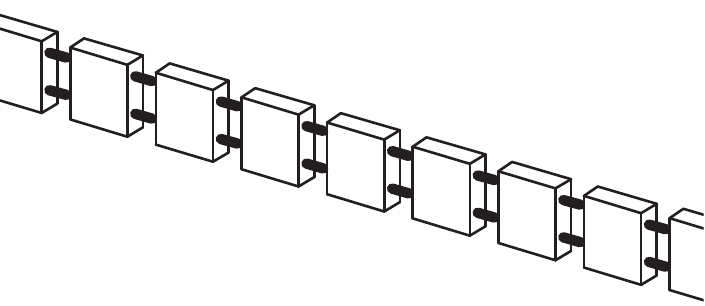}
	\caption{Basic building block for $L=1$ and its infinite interconnection.}
	 \label{fig:1DInfinite}
\end{figure}

\begin{figure}[!t]
	\centering
	\includegraphics[height=1.8cm]{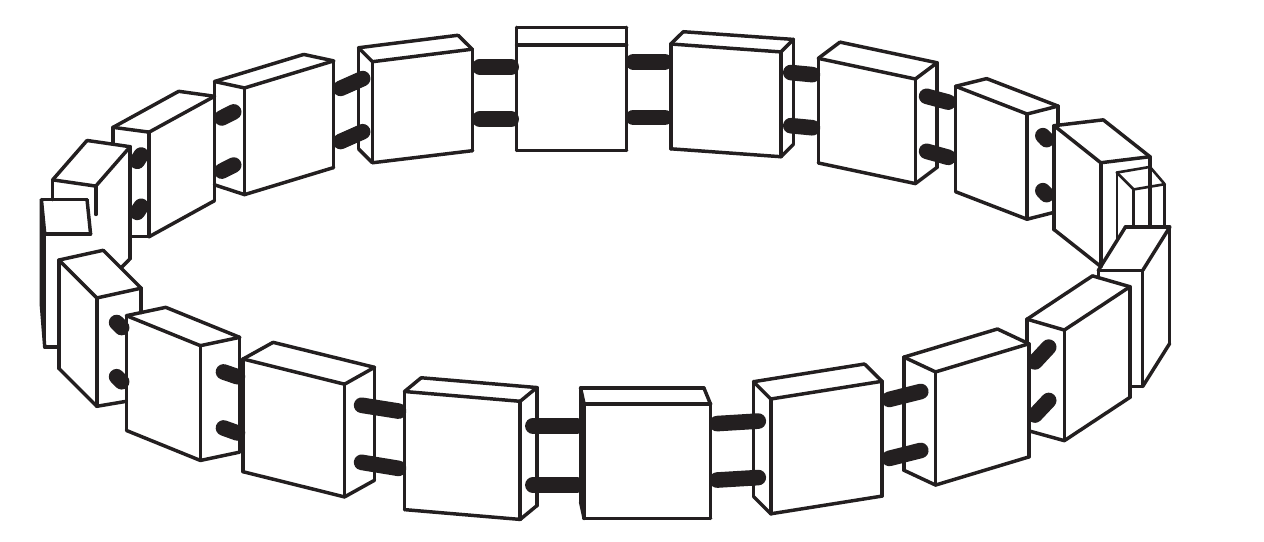}    
	\caption{Periodic interconnection for $L=1$.}  
	\label{fig:1DPeriodic}                                 
\end{figure}

\begin{figure}[!t]
	\centering
	\includegraphics[height=1.8cm]{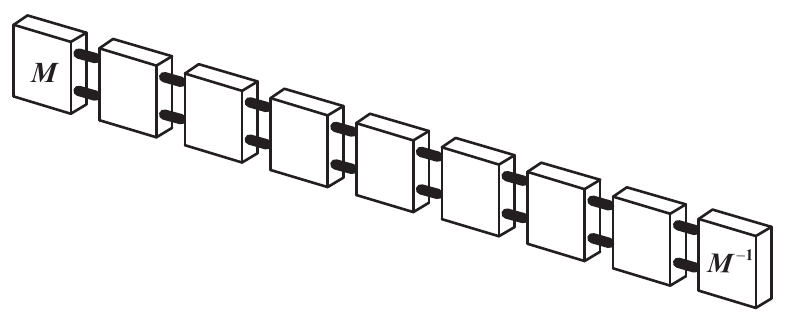}    
	\caption{Finite extent system for $L=1$ with boundary condition matrix $M$.}  
	\label{fig:1DFinite}                                 
\end{figure}

For a schematic illustration, a basic building block with $L=1$, and its corresponding three types of interconnections are depicted in Fig.~\ref{fig:1DInfinite}-Fig.~\ref{fig:1DFinite}.

As explored in \cite{DAndrea2003,Langbort2005}, all interconnected systems of interest can be captured by the following abstract differential equation on the Hilbert space $\ell_2$:
\begin{equation}\label{Sigma2}
\begin{aligned}
\begin{bmatrix}
\frac{\partial x(t)}{\partial t}\\  \Delta v(t) 
\end{bmatrix}&=\begin{bmatrix}
A_{TT} & A_{TS}\\A_{ST} & A_{SS}
\end{bmatrix}\begin{bmatrix}
x(t) \\ v(t)
\end{bmatrix},\\
x(0)&=x_0\in \ell_2,
\end{aligned} 
\end{equation}
where 
\begin{equation}
\Delta = \begin{bmatrix}
\Delta_1 & & \\
& \ddots & \\
& & \Delta_{L}
\end{bmatrix},
\end{equation}
and $\Delta_{i}$ denotes structured operator in the $i$-th
 spatial dimension: \begin{equation}
\Delta_{i}=\begin{cases}
diag\left(\mathbf{S}_iI_{n_i},\mathbf{S}^{-1}_iI_{n_{-i}} \right),~\text{for infinite}\\
\text{or periodic spatial extent,}\\
\mathcal{C}_i,~\text{for finite spatial extent,}
\end{cases}
\end{equation}
where $\mathbf{S}_i$ and $\mathcal{C}_i$ share the same mechanisms with operators defined in \cite{Langbort2005}. 

The system (\ref{Sigma2}) is said to be well-posed if the bounded linear operator $\Delta - A_{SS}$ is invertible on $\ell_2$. The well-posedness can be interpreted as the existence and uniqueness of solution. It is assumed that system (\ref{Sigma2}) is well-posed in the rest of this paper since it would make more practical sense to analysis the stability  in the presence of wellposedness.

\begin{assumption}
	$\Delta-A_{SS}$ is invertible on $\ell_2$. 
\end{assumption}

A well-posed system (\ref{Sigma2}) has a unique solution for any $x_0\in \ell_2$:
\begin{equation}
x(t) = exp(\mathbf{A}t)x_0,
\end{equation}
where $\mathbf{A} := A_{TT} + A_{TS}(\Delta-A_{SS})^{-1}A_{ST}$ is a bounded operator on $\ell_2$ and generates a strong continuous semigroup $exp(\mathbf{A}t)$ formed by \begin{equation}
exp(\mathbf{A}t):=\sum_{n=0}^{\infty}\frac{(\mathbf{A}t)^n}{n!}.
\end{equation}

Under the assumption of well-posedness, the exponential stability of $\Sigma$ is equivalent to that of system (\ref{Sigma2}), i.e, there exist $\alpha>0$, $\beta>0$,  such that the continuous semigroup $exp(\mathbf{A}t)$ satisfies \begin{equation}\label{estable}
\|exp(\mathbf{A}t)\|_{\ell_2}\leq \alpha exp(-\beta t),~\forall t\in \mathbb{R}^{+}.
\end{equation}

The problem addressed here is to study the exponential stability of $\Sigma$ by resorting to sum-of-squares (SOS) technique for trigonometric polynomials. 

\subsection{Trigonometric polynomials}
To derive the main results of this paper, some concepts relevant to trigonometric polynomials are presented in the sequel.

Let $\mathbf{z}=(z_1,...,z_L)$ be the $L$-dimensional complex variables defined on the unit $L$-circle $\mathbb{T}^L=\left\{ \mathbf{z}\in \mathbb{C}^L :~|z_i|=1,i=1,...,L \right\}$, $\mathbf{z}^{\mathbf{d}}=z^{d_1}_1 \cdots z^{d_L}_L $ is a $L$-variate monomial of degree $\mathbf{d}=\left[ d_1,\cdots, d_L \right] \in \mathbb{Z}^L$. 

In this paper, we deal with the real-valued (Hermitian) trigonometric polynomial
\begin{equation}\label{RP}
F(\mathbf{z})  = \sum^{\mathbf{n}}_{\mathbf{d}=-\mathbf{n}}f(\mathbf{d})\mathbf{z}^{\mathbf{d}},~f(-\mathbf{d}) = f^*(\mathbf{d}),
\end{equation}
where $\mathbf{n}=(\mathbf{n}_1,...,\mathbf{n}_L)\in \mathbb{Z}^L$ gathers the maximum degree for each variable $z_i$, regarded as the degree of $F(\mathbf{z})$. The set of real-valued trigonometric polynomials  is denoted by $\mathbb{RP}[\mathbb{T}^L]$.

 \begin{definition}\cite{Dumitrescu2007}
 	A trigonometric polynomial $F\in \mathbb{RP}[\mathbb{T}^L]$ is said to be sum-of-squares (SOS), if it can be written as
 	\begin{equation}
 	F(\mathbf{z}) = \sum^r_{l=1}V_{l}(\mathbf{z})V^*_l(\mathbf{z}^{-1}), 
 	\end{equation} 
 	where $V_l(\mathbf{z})$ are positive orthant polynomials, i.e, only monomials $\mathbf{z}^{\mathbf{d}}$ with $\mathbf{d}\geq 0 $ are contained, and $V^*_l(\mathbf{z})$ denotes the polynomial with complex conjugated coefficients. 
 \end{definition}

It is obvious that any SOS polynomial is nonnegative on the unit $L$-circle, and an important theoretical result is that any polynomial (\ref{RP}) positive on the unit $L$-circle is SOS, see \cite{Dumitrescu2007} for details. 

Moreover, consider the set 
\begin{equation}
\mathcal{D} = \left\{ \mathbf{z}\in \mathbb{T}^L: D_i(\mathbf{z})\geq 0, i=1:\mathcal{V} \right\},
\end{equation}
defined by the positivity of some given trigonometric polynomials $D_i(\mathbf{z})$,  $i=1:\mathcal{V}$, the following lemma characterize the trigonometric polynomials that are positive on $\mathcal{D}$.

\begin{lma} \cite{Dumitrescu2006} \label{lma1}
	If a polynomial $F\in \mathbb{RP}[\mathbb{T}^L]$ is positive on $\mathcal{D}$, then there exist SOS polynomials $H_i(\mathbf{z})$, $i=0:\mathcal{V}$, such that 
	\begin{equation}
	F(\mathbf{z}) = H_0(\mathbf{z}) + \sum_{i=1}^{\mathcal{V}} D_i(\mathbf{z}) H_i(\mathbf{z}).
	\end{equation}
\end{lma}

\section{Main results} \label{sec3}

\subsection{Full Infinite Interconnections}

In this section, the spatial interconnections are restricted to be infinite extent for each spatial direction. It follows from the results of \cite{Bamieh2002} that the stability of solution for system (\ref{Sigma1}) can be equivalently checked by looking at its corresponding Fourier-transformed form, i.e, the system $\Sigma$ with infinite spatial interconnections is exponentially stable if and only if $A(\mathbf{z})$ is Hurwitz for all $\mathbf{z}\in \mathbb{T}^L$ (i.e, all its eigenvalues have negative real parts), where 
\begin{align}\label{hatA}
A(\mathbf{z})&=A_{TT}+A_{TS}(\Delta(\mathbf{z})-A_{SS})^{-1}A_{ST},\\
\Delta(\mathbf{z}) &= diag\left\{\left.\begin{bmatrix}
z_iI_{n_i} & \\ & z^{-1}_{i}I_{n_{-i}}
\end{bmatrix}\right|_{i=1}^{L}\right\}.
\end{align}
Note that $\Delta(\mathbf{z})-A_{SS}$ is invertible for all $\mathbf{z}\in \mathbb{T}^L$ due to the assumption of well-posedness (see \cite{Langbort2005} for details). 

Inspired by the idea of rational parameterization \cite{Chesi2019}, let us define 
\begin{equation}
h(\mathbf{z}) = det(\Delta(\mathbf{z}) - A_{SS})
\end{equation}
and $H(\mathbf{z})$ is the matrix polynomial 
\begin{equation}
H(\mathbf{z}) = A_{TT}h(\mathbf{z}) + A_{TS}adj\left( \Delta(\bm{z}) - A_{SS} \right)A_{ST},
\end{equation}
then $A(\mathbf{z})$ can be expressed as
\begin{equation}\label{AHh}
A(\mathbf{z}) = \frac{H(\mathbf{z})}{h(\mathbf{z})}.
\end{equation}

 The following theorem provides a necessary and sufficient condition for establishing the stability of interest.
\begin{theorem}\label{thm1}
	The complex matrix $A(\mathbf{z})$ is Hurwitz over $\mathbb{T}^L$ if and only if 
	\begin{enumerate}[(i)]
		\item  \label{c1} $A(\mathbf{z})|_{\mathbf{z}=\mathbf{1}_L}$ is Hurwitz;
		\item 	\label{c2} $F(\mathbf{z}):=det(-W(\mathbf{z}))$ is positive on the unit $L$-circle $\mathbb{T}^L$, where 
		\begin{align}
		W(\mathbf{z})& = K(\mathbf{z}) \otimes I_{n_0} + I_{n_0} \otimes  \overline{K(\mathbf{z})},\\
		K(\mathbf{z})&= H(\mathbf{z})\overline{h(\mathbf{z})}.
		\end{align}
	\end{enumerate}
\end{theorem}
\emph{\textbf{Proof.}
		``$\Rightarrow$" Suppose that $A(\mathbf{z})$ is Hurwitz over $\mathbb{T}^L$, it is obvious that $A(\mathbf{z})|_{\mathbf{z}=\mathbf{1}_L}$ is Hurwitz. 
		Let \begin{equation}
		\hat{W}(\mathbf{z}) = A(\mathbf{z})\otimes I_{n_0}+I_{n_0}\otimes \overline{A(\mathbf{z})},
		\end{equation}
			then it follows from the properties of the Kronecker product that
		\begin{equation}
		spec(\hat{W}(\mathbf{z})) = \left\{ \lambda_k(\mathbf{z}) + \overline{\lambda_l(\mathbf{z})}, k,l=1,...,n_0 \right\}
		\end{equation}
				where $\lambda_i(\mathbf{z})$, $i=1,...,n_0$ denotes the $i$-th eigenvalue of $A(\mathbf{z})$, thus $\hat{W}(\mathbf{z})$ is Hurwitz for all $\mathbf{z}\in \mathbb{T}^L$.  It is observed that the eigenvalues of $\hat{W}(\mathbf{z})$ are symmetric with respect to the real axis,  which implies \begin{equation}
				\hat{F}(\mathbf{z}) = det(-\hat{W}(\mathbf{z}))
		\end{equation}
		is  positive over $\mathbb{T}^L$. Thus, 
		\begin{equation}\label{FhatF}
		F(\mathbf{z})=det\left(-W(\mathbf{z})\right) = |h(\mathbf{z})|^{2n^2_0} \hat{F}(\mathbf{z})
		\end{equation}
		is positive.}
		
		\emph{``$\Leftarrow$" Suppose that $A(\mathbf{z})|_{\mathbf{z}=\mathbf{1}_L}$ is Hurwitz, and $F(\mathbf{z})$ is positive on the unit $L$-circle. It follows from (\ref{FhatF}) that $\hat{F}(\mathbf{z})$ is positive. Assume that there exists $\mathbf{z}_1\in \mathbb{T}^L$ such that $A(\mathbf{z})|_{\mathbf{z}=\mathbf{z}_1}$ is not Hurwitz. Due to $A(\mathbf{z})|_{\mathbf{z}=\mathbf{1}_L}$ is Hurwitz, there exists $\mathbf{z}_2\in \mathbb{T}^L$ such that $A(\mathbf{z})|_{\mathbf{z}=\mathbf{z}_2}$ has some eigenvalues with null real part according to the continuity of the eigenvalues of $A(\mathbf{z})$, which implies that $\hat{W}(\mathbf{z})|_{\mathbf{z}=\mathbf{z}_2}$ is singular since if $\lambda$ is an eigenvalue of $A(\mathbf{z})|_{\mathbf{z}=\mathbf{z}_2}$ with null real part,  $\lambda + \overline{\lambda}=0$ is an eigenvalue of $\hat{W}(\mathbf{z})|_{\mathbf{z}=\mathbf{z}_2}$.  Thus, $$\hat{F}(\mathbf{z})|_{\mathbf{z}=\mathbf{z}_2}=det\left(-\hat{W}(\mathbf{z})|_{\mathbf{z}=\mathbf{z}_2}\right) =0,$$ which contradicts the positiveness of $\hat{F}(\mathbf{z})$.  The proof is  completed. }  $\qedd$

Theorem \ref{thm1} shows that the stability of system $\Sigma$ with infinite interconnections can be equivalently checked via two sub-conditions. The former is readily tested, while the latter requires to check the positivity of a polynomial $F(\mathbf{z})$ over $\mathbb{T}^L$. 

Note that $F(\mathbf{z})$ is a real-valued trigonometric polynomial, $F(\mathbf{z})$ is positive over $\mathbb{T}^L$ if and only if, there exists $\epsilon>0$ such that
\begin{equation}\label{PositiveRelaxed}
F(\mathbf{z}) - \epsilon >0, ~\forall \mathbf{z} \in \mathbb{T}^L. 
\end{equation}
From the previous discussion, it follows that $F(\mathbf{z})>0$ if and only if,  there exists $\epsilon>0$ such that
\begin{equation}
F(\mathbf{z}) - \epsilon \text{ is SOS. }
\end{equation}

\subsection{Mixed  Interconnections}\label{Sec3_2}
In this section, we present the counterpart of previous result for mixed infinite-periodic interconnected system, and its extensions on spatially reversible finite extent systems.  

Consider the system $\Sigma$ in (\ref{Sigma1}) with 
\begin{equation}
\mathbb{D}_i=\begin{cases}
\mathbb{Z},~i=1,...,l,\\
\mathbb{Z}_{N_i}+1,i=l+1,...,L,
\end{cases}
\end{equation}
which indicates that the first $l$ spatial coordinates are infinite interconnected, and the last $L-l$ are periodic. If we define
\begin{equation}
\mathbb{S}_{i} = \left\{z\in \mathbb{T}:~z^{N_i}=1 \right\},~i=l+1,...,L,
\end{equation}
the system $\Sigma$ with this type of interconnections is exponentially stable if and only if 
\begin{equation}\label{PHurwitz}
A(\mathbf{z})~\text{is Hurwitz,} ~\forall \mathbf{z}\in \mathbb{G}:=\mathbb{T}^{l} \times \mathbb{S}_{l+1} \times \cdots \times \mathbb{S}_{L}. 
\end{equation}

Obviously, if the two sub-conditions derived in Theorem 1 are satisfied, (\ref{PHurwitz}) holds since $\mathbb{G}\subset \mathbb{T}^L$; however, to the best of our knowledge, the reciprocal may not be true. 

The idea of robust stabilizability functions in \cite{Chesi2014} encourages us to find another trigonometric polynomial, which is positive on $\mathbb{G}$ if and only if (\ref{PHurwitz}) holds.

Let us define
\begin{equation}
K(\mathbf{z}) = H(\mathbf{z})\overline{h(\mathbf{z})}=|h(\mathbf{z})|^2 A(\mathbf{z}),
\end{equation}
and its characteristic polynomial
\begin{equation}
m(\lambda,\mathbf{z}) = det(\lambda I_{n_0} - K(\mathbf{z})).
\end{equation}

Note that $m(\lambda,\mathbf{z}) $ can be expressed as 
\begin{equation}
m(\lambda,\mathbf{z}) = \sum^{n_0}_{i=0} m_i(\mathbf{z})\lambda^i,
\end{equation}
where $m_i(\mathbf{z})$ are trigonometric polynomials. Furthermore, we define
\begin{equation}
m_{conj}(\lambda,\mathbf{z}) = \sum_{i=0}^{n_0} \overline{m_i(\mathbf{z})} \lambda^i
\end{equation}
and 
\begin{equation} \label{varPhi}
\varphi(\lambda,\mathbf{z}) = m(\lambda,\mathbf{z})m_{conj}(\lambda,\mathbf{z}),
\end{equation}
then $\varphi(\lambda,\mathbf{z})$ can be expressed by
\begin{equation}
\varphi(\lambda,\mathbf{z}) = \sum_{i=0}^{2n_0}\varphi_i(\mathbf{z})\lambda^i,
\end{equation}
in which $\varphi_i(\mathbf{z})$ are real-valued trigonometric polynomials.

It can be found that the roots of $m(\lambda,\mathbf{z})$ and  $m_{conj}(\lambda,z)$ share the same real parts, thus the stability of interest can be  established by investigating stability of the roots of $\varphi(\lambda,\mathbf{z})$ in (\ref{varPhi}).

The Routh-Hurwitz table of $\varphi(\lambda,\mathbf{z})$ can be written as
\begin{equation}
\begin{cases}
e_{0,j}(\mathbf{z})= \varphi_{2n_0-2j}(\mathbf{z}),~\forall j=0,...,n_0,\\
e_{1,j}(\mathbf{z})= \varphi_{2n_0-2j-1}(\mathbf{z}),~\forall j=0,...,n_0-1,
\end{cases}
\end{equation}
and
\begin{equation}
\begin{aligned}
e_{i,j}(\mathbf{z})=\left \llbracket \begin{array}{cc}
e_{i-2,0}(\mathbf{z}) & e_{i-2,j+1}(\mathbf{z})\\
e_{i-1,0}(\mathbf{z}) & e_{i-1,j+1}(\mathbf{z})
\end{array}\right \rrbracket,~\forall i&= 2,...,2n_0,\\
j&=0,1,...
\end{aligned}
\end{equation}
Construct $\bar{e}_{i,0}(\mathbf{z})$ and $\hat{e}_{i,0}(\mathbf{z})$ from the obtained $e_{i,j}(\mathbf{z})$ by 
\begin{equation}\label{routh}
\begin{aligned}
\bar{e}_{0,0}(\mathbf{z}) &= e_{0,0}(\mathbf{z}),~\hat{e}_{0,0}(\mathbf{z})=1,\\
\frac{\bar{e}_{i,0}(\mathbf{z})}{\hat{e}_{i,0}(\mathbf{z})}&= e_{i,0}(\mathbf{z}),~\hat{e}_{i,0}(\mathbf{z})=\prod_{k=i-1,i-3,...}\bar{e}_{k,0}(\mathbf{z}),~\forall i=1,...,2n_0.
\end{aligned}
\end{equation}
Then, define the set \begin{equation}
\mathcal{N}=\left\{  i=0,...,n_r: \bar{e}_{i,0}(\mathbf{z}) \text{ is a non-positive constant} \right\}
\end{equation} and let $F_k(\mathbf{z})$, $k=1,...,n_f$, be the non-constant polynomials among $\bar{e}_{i,0}(\mathbf{z}),~i=0,...,2n_0$.

\begin{theorem}\label{thm2}
	The complex matrix $A(\mathbf{z})$ is Hurwitz on $ \mathbb{G}$ if and only if the following two sub-conditions are satisfied
\begin{enumerate}[(i)]
	\item $\mathcal{N}=\emptyset$;
	\item $F_k(\mathbf{z})$ ($k=1,...,n_f$) is positive on $\mathbb{G}$.
\end{enumerate}
\end{theorem}
\emph{\textbf{Proof.}
	``$\Rightarrow$" Suppose that for each $\mathbf{z}\in \mathbb{G}$, $A(\mathbf{z})$ is Hurwitz, it follows from the Routh-Huwritz  criterion that $\mathcal{N}=\emptyset$, and \begin{equation}
	F_k(\mathbf{z})>0,~\forall k=1,...,n_f,~\forall \mathbf{z}\in \mathbb{G}.
	\end{equation}}
	 \emph{
	``$\Leftarrow$" Suppose that $\mathcal{N}=\emptyset$, and $F_k(\mathbf{z})>0$ $(i=1,...,n_f)$ for all $\mathbf{z}\in \mathbb{G}$. This implies that
	\begin{equation}
	\bar{e}_{i,0}>0,~\forall i=0,...,2n_0,~\forall \mathbf{z}\in \mathbb{G}.
	\end{equation}
	Hence, 
	\begin{equation}
	K(\mathbf{z}) \text{ is Hurwitz } \forall \mathbf{z}\in \mathbb{G},
	\end{equation}
	Note that $h(\mathbf{z})\neq 0$ due to the assumption of well-posedness, thus
	\begin{equation}
	A(\mathbf{z}) = \frac{K(\mathbf{z})}{|h(\mathbf{z})|^2} \text{ is Hurwitz } \forall \mathbf{z}\in \mathbb{G}.
	\end{equation} 
	The proof is completed.} $\qedd$

If we define 
\begin{equation}
F(\mathbf{z}) = \min_{k=1,...,n_f} F_k(\mathbf{z}),~\forall \mathbf{z}\in \mathbb{G},
\end{equation}
the second sub-condition in Theorem \ref{thm2} will amount to the positivity of polynomial $F(\mathbf{z})$ over $\mathbb{G}$. However, it is generally hard to obtain an exact expression for such $F(\mathbf{z})$. Thus, we prefer to check the positivity of $F_k(\mathbf{z})$ over  $\mathbb{G}$.

It is observed that
\begin{equation}
\mathbb{G} = \left\{ \mathbf{z}\in \mathbb{T}^L :  D_i(\mathbf{z})\geq 0,~i=1,...,L-l  \right\},
\end{equation}
in which
\begin{equation}
D_i(\mathbf{z}) = z^{N_{i+l}}_{i+l} - 1 = D_{i}(\mathbf{d}) \mathbf{z}^{\mathbf{d}},
\end{equation}
 thus it follows from Lemma \ref{lma1} that the system $\Sigma$ considered in this section is exponentially stable if and only if $\mathcal{N}=\emptyset$, and for each $k=1,...,n_f$, there exist $\epsilon>0$, and SOS polynomials $H_{k,i}(\mathbf{z})$, $i=0:L-l$, 
	such that 
	\begin{equation}\label{sos2}
	F_k(\mathbf{z})  - \epsilon = H_{k,0}(\mathbf{z}) +  \sum_{i=1}^{L-l}D_i(\mathbf{z})H_{k,i}(\mathbf{z}). 
	\end{equation}

In what follows, we turn our attention to the spatially reversible finite extent systems (see \cite{Langbort2005} for details). The $i$-th spatial index $k_i$ is restricted in the following set
\begin{equation}
\mathbb{D}_i = \begin{cases}
\mathbb{Z}_{N_i}+1,~i=1,...,l,\\
\left\{ 1,...,N_i \right\},~i=l+1,...,L,
\end{cases}
\end{equation}
i.e., the first $l$ spatial coordinates are periodic interconnected, and the last $L-l$ are spatially reversible finite extent.

Benefiting from the results derived in \cite{Langbort2005}, the stability of such system can be checked by looking at the corresponding periodic system indexed over the set
\begin{equation}
\mathbb{G} := 
\mathbb{Z}_{N_1}\times \cdots \times \mathbb{Z}_{N_l} \times \mathbb{Z}_{2N_{l+1}}\times \cdots \times \mathbb{Z}_{2N_{L}}.
\end{equation}

Thus, the stability of interest can be established by following the same pattern shown above. The only difference is that the result pertaining to the corresponding periodic system only yield sufficient condition for finite extent case.

\subsection{Generalized Trace Parameterization}

In this section, the generalized trace parameterization of trigonometric polynomial is recalled \cite{Dumitrescu2007}, so that the theoretical results derived in the previous subsections can be quantified by two semidefinite programs (SDPs).

For any  polynomial $F(\mathbf{z})$ defined in (\ref{RP}), let
\begin{equation}
p(z_i) = \begin{bmatrix}
1 & z_i & \cdots & z^{\hat{\mathbf{n}}_i}_i
\end{bmatrix}^T,~i=1:L,
\end{equation}
to be the vector that gathers the canonical basis for polynomials of degree $\hat{\mathbf{n}}_i$ in   $i$-th variable $z_i$, where $\hat{\mathbf{n}}=(\hat{\mathbf{n}}_1,...,\hat{\mathbf{n}}_L)\geq \mathbf{n}$, and the vector
\begin{equation}
p(\mathbf{z}) = p(z_L) \otimes  \cdots \otimes p(z_1),
\end{equation}
of length $\hat{N} = \Pi_{i=1}^{L}(\hat{\mathbf{n}}_i+1)$ to be the canonical basis for $L$-variate polynomials of degree $\hat{\mathbf{n}}$, where $\otimes$ represents the Kronecker product. Then,  there exist Gram matrix parameterization for $F(\mathbf{z})$, i.e, 
\begin{equation}\label{Gram}
F(\mathbf{z}) = p^T(\mathbf{z}^{-1})\cdot G \cdot p(\mathbf{z}).
\end{equation}
where $G\in \mathbb{C}^{\hat{N}\times \hat{N}}$ is a Hermitian matrix, called a Gram matrix associated with $F(\mathbf{z})$, the set of Gram matrices associated with $F(\mathbf{z})$ is denoted by $\mathcal{G}(F)$.

The general relation between the coefficients of the polynomial $F(\mathbf{z})$ and its Gram matrix is given by the following lemma.
\begin{lma}\cite{Dumitrescu2007}
	For a trigonometric polynomial $F$ defined in (\ref{RP}), and $G\in \mathcal{G}(F)$, then the relation
	\begin{equation}\label{TrGram}
	tr[T(\mathbf{d})\cdot G] = \begin{cases}
	f(\mathbf{d}),~\mathbf{d} \in [-\mathbf{n},\mathbf{n}]\\
	0,~\mathbf{d} \in [-\hat{\mathbf{n}},\hat{\mathbf{n}}]\backslash [-\mathbf{n},\mathbf{n}]
	\end{cases}
	\end{equation}
	holds, where
	\begin{equation}\label{Toeplitz}
	T(\mathbf{d}) = T_L(d_L)\otimes \cdots \otimes T_1(d_1),
	\end{equation}
	and $T_i(d_i)\in \mathbb{R}^{(\hat{\mathbf{n}}_i+1)\times (\hat{\mathbf{n}}_i+1)}$ are elementary Toeplitz matrices with ones only on the $d_i$-th diagonal. 
\end{lma}

The relation (\ref{TrGram}) is called the generalized trace parameterization.
It is found that the size of Gram matrix $G$ is defined by the $L$-tuple $\hat{\mathbf{n}}$ that assumed to be unknown previously, which leads to an important result on trigonometric polynomial that any polynomial $F$ in (\ref{RP}) is SOS if and only if there exists a semipositive definite Gram matrix $G$ associated with $F(\mathbf{z})$, i.e, there exist an $L$-tuple $\hat{\mathbf{n}}$, and matrix $G\in \mathbb{C}^{\hat{N}\times \hat{N}}\geq 0$, such that (\ref{TrGram}) holds. 

With the generalized trace parameterization, the derived theoretical results can be quantified by semi-definite programs in optimization problem. Specifically, for the globally positive trigonometric polynomial $F(\mathbf{z})$ from 2) in Theorem \ref{thm1}, let us define
\begin{equation}\label{opt1}
\begin{aligned}
\epsilon^*(\bm{e}) &= \sup_{\epsilon,~G}\epsilon\\
s.t.~&\begin{cases}
g(\mathbf{d}) = tr[ T(\mathbf{d}) G ],\\
\mathbf{n}_G = \mathbf{n}_F+\bm{e},\\
G\in \mathbb{C}^{N_G\times N_G}\geq 0,\\
\end{cases}
\end{aligned}
\end{equation}
where $\mathbf{n}_F$ denotes the degree of $F(\mathbf{z})$, 
$\bm{e}$ has nonnegative elements, and 
\begin{equation}
g(\mathbf{d}) = 
\begin{cases}
f(\mathbf{d}),~\mathbf{d} \in [-\mathbf{n}_F,\mathbf{n}_F]\\
0,~\mathbf{d} \in [-\mathbf{n}_G,\mathbf{n}_G] \backslash [-\mathbf{n}_F,\mathbf{n}_F]
\end{cases}
\end{equation}
in which, $f(\mathbf{d})$ denotes the symmetric representation (\ref{RP}) for $F(\mathbf{z}) - \epsilon$.
Then, 2) holds if and only if $\epsilon^*(\bm{e})> 0$ for some nonnegative $\bm{e}$.

For the positivity of $F_k(\mathbf{z})$ from 2) in Theorem \ref{thm2} on domain $\mathbb{G}$, let us define 
\begin{equation}\label{opt2}
\begin{aligned}
\epsilon^*(\bm{e}) &= \sup_{\epsilon,~G_{k,i}}\epsilon\\
s.t.&~\forall k=1,...,n_f\\
&\begin{cases}
	g_k(\mathbf{d}) -\sum^{L-l}_{i=1} \varphi_{k,i}(\mathbf{d})= tr\left[ T_{k,0}(\mathbf{d})\cdot G_{k,0} \right],\\
\mathbf{n}_{G_{k,0}} = 2(\mathbf{n}_k+\bm{e}),\\
\mathbf{n}_{k} = \lceil \frac{1}{2} \max \left\{ \mathbf{n}_{F_k}, \mathbf{n}_{D_1},...,\mathbf{n}_{D_{L-l}} \right\}\rceil, \\
\mathbf{n}_{G_{k,i}} = \mathbf{n}_{G_{k,0}}  - \mathbf{n}_{D_i},~i=1:L-l  \\
G_{k,i}\in \mathbb{C}^{N_{G_{k,i}}\times N_{G_{k,i}}}\geq 0,~i=0:L-l\\
\end{cases}
\end{aligned}
\end{equation}
where $\mathbf{n}_{F_k}$ denotes the degree of $F_k(\mathbf{z})$, 
$\bm{e}$ has nonnegative elements, and 
\begin{align}
N_{G_{k,i}} &= \Pi_{v=1}^L({\mathbf{n}}^v_{G_{k,i}}+1),~\varphi_{k,i}(\mathbf{d}) = \sum_{\mathbf{d_1}+\mathbf{d_2}=\mathbf{d}}f^{i}_d(\mathbf{d}_1)f^{k,i}_h(\mathbf{d}_2),\\
\sum_{\mathbf{d}=-\mathbf{n}_{D_i}}^{\mathbf{n}_{D_i}}f^i_d(\mathbf{d})\mathbf{z}^{\mathbf{d}}&:=D_i(\mathbf{z}),~f^{k,i}_h(\mathbf{d}):=tr\left[ T_{k,i}(\mathbf{d})\cdot G_{k,i} \right],\\
g_k(\mathbf{d})& = 
\begin{cases}
f_k(\mathbf{d}),~\mathbf{d} \in [-\mathbf{n}_{F_k},\mathbf{n}_{F,k}]\\
0,~\mathbf{d} \in [-\mathbf{n}_{G_{k,0}},\mathbf{n}_{G_{k,0}}] \backslash [-\mathbf{n}_{F_k},\mathbf{n}_{F_k}]
\end{cases},\\
\sum_{\mathbf{d}=-\mathbf{n}_{F_k}}^{\mathbf{n}_{F_k}}f_k(\mathbf{d})\mathbf{z}^{\mathbf{d}}&:=F_k(\mathbf{z}) - \epsilon,
\end{align}
in which $T_{k,i}(\mathbf{d})$ ($i=0:L-l$) are defined by (\ref{Toeplitz}) associated with $\mathbf{n}_{G_{k,i}}$. 
We determine that 2) holds, if $\epsilon^*(\bm{e})> 0$ in SDP \eqref{opt2} for some nonnegative $\bm{e}$. This test gives no false positive, and in the example section we will see that if 2) holds, a positive $\epsilon$ can be readily found in general.

\section{Examples}\label{sec4}
In this section, several examples are given to demonstrate the efficiency of the derived theoretical results. 

\textbf{Example 1: } Let us consider the problem of determining whether an infinite interconnected system $\Sigma$ is exponentially stable with $L=2$, where   
\begin{equation}
\begin{cases}
\begin{aligned}
A_{TT} &= \begin{bmatrix}
-0.5 & 0\\ 0 & -1
\end{bmatrix},~A_{TS} = \begin{bmatrix}
1 & 0 & 0 & 2\\ 0 & 0 & 0.5 & 0
\end{bmatrix},\\
A_{ST} &=\begin{bmatrix}
0 & 0.5 \\ 1 & 0 \\ -0.5 & 0\\ 0 & 0
\end{bmatrix},~A_{SS}=\begin{bmatrix}
0 & 0 & 0 & 0\\0 & 0 & 0 & 0\\0 & 0 & 0 & 0\\0 & 0 & 0 & 0
\end{bmatrix}.
\end{aligned}
\end{cases}
\end{equation}

Firstly, it is readily verified that the constant matrix 
\begin{equation}
\begin{aligned}
\mathbf{A}(\mathbf{z})|_{z_1=1,z_2=1}&=A_{TT} + A_{TS}(I - A_{SS})^{-1}A_{ST}\\
&=\begin{bmatrix}
-0.5 & 0.5\\
-0.25 & -1.0
\end{bmatrix}
\end{aligned}
\end{equation}
has all eigenvalues with negative real part, i.e, the first sub-condition in Theorem \ref{thm1} holds. Thus, we turn our attention to the second sub-condition. 
 
The polynomial $h(\mathbf{z})$ is given by
\begin{equation}
h(\mathbf{z}) = det\left( \Delta(\mathbf{z}) - A_{SS} \right) = 1,
\end{equation}
and the matrix polynomial $H(\mathbf{z})$ 
\begin{equation}
\begin{aligned}
H(\mathbf{z})& = A_{TT}h(\mathbf{z}) + A_{TS}adj(\Delta(\mathbf{z}) - A_{SS})A_{ST} \\
& = \begin{bmatrix}
-0.5 & 0.5z^{-1}_1 \\
-0.25 z^{-1}_2 & -1
\end{bmatrix}.
\end{aligned}
\end{equation}

It is straightforward to figure out  
\begin{equation}
\begin{aligned}
F(\mathbf{z}) &= 0.015625z^{-2}_1z^{-2}_2 + 0.5625 z^{-1}_1z^{-1}_2 + 4.46875\\
&+0.5625z_1z_2 + 0.015625z^{2}_1z^2_2,
\end{aligned}
\end{equation}
then using the YALMIP toolbox to solve the optimization problem (\ref{opt1}) for $\bm{e} = (0,0)$, it is found that 
\begin{equation*}
\epsilon = 3.3750>0.
\end{equation*}
Thus, it can be concluded from Theorem \ref{thm1} that the system $\Sigma$ is exponentially stable. 

\begin{figure}
\centering
		\includegraphics[height=6cm]{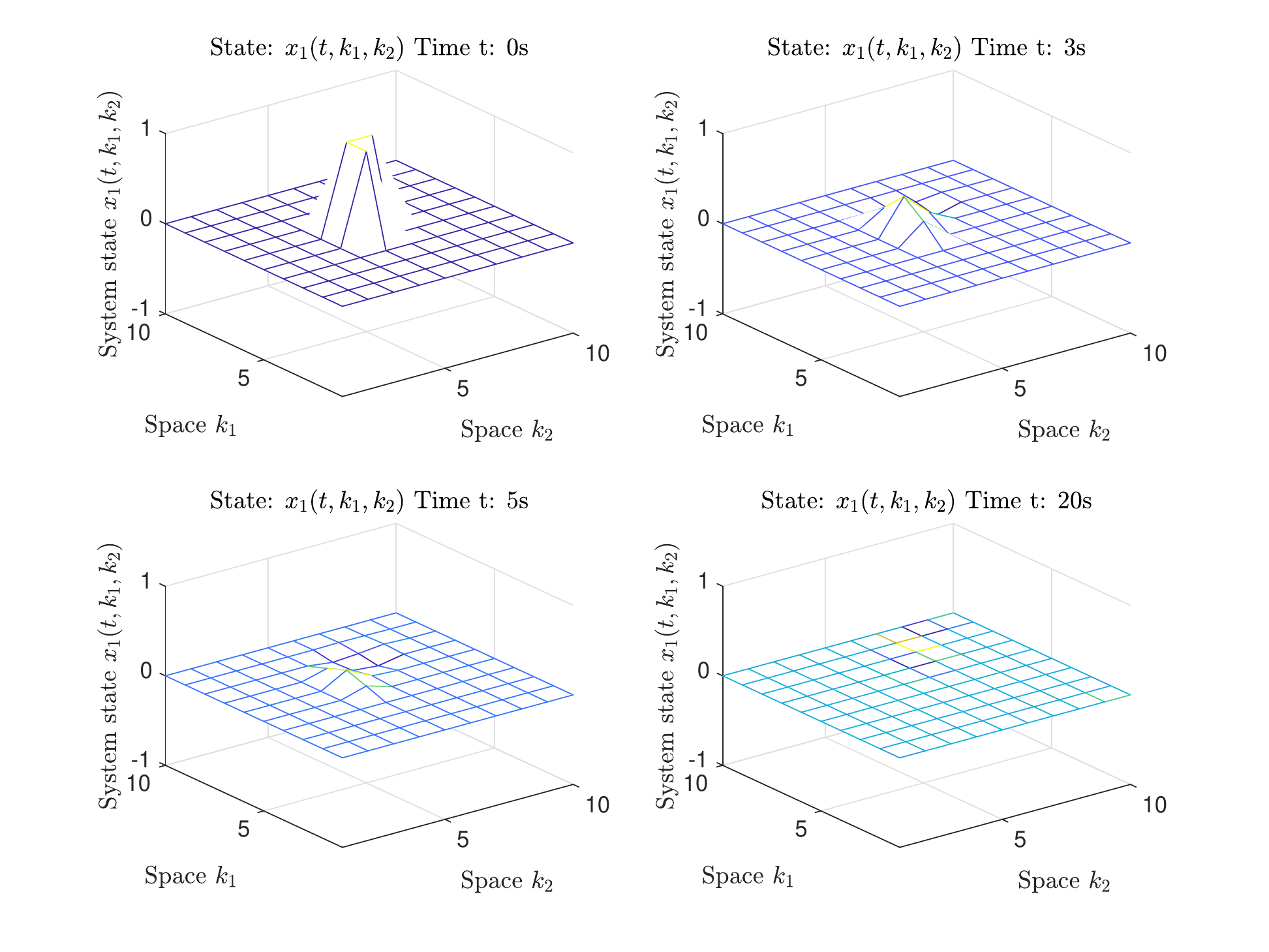}    
		\caption{State response $x_{1}(t,k_1,k_2)$ at different time}  
		\label{fig:ex1x1}                                 
\end{figure}

The multidimensional (MD) toolbox is used to simulate the motion of this system under the following initial condition:
\begin{equation}\label{initC}
\begin{aligned}
\forall i=1,2,~x_i(0,k_1,k_2) = \begin{cases}
1,~&\text{if } k_1=5,~k_2=5; \\
&\quad k_1=6,~k_2=5; \\
&\quad k_1=6,~k_2=6,\\
0,~&\text{otherwise.}
\end{cases}
\end{aligned}
\end{equation}
The state response $x_1(t,k_1,k_2)$ at different time ($t=0$, $3$, $5$, $20$s) is shown in 
Fig.~\ref{fig:ex1x1}, which illustrates the stability of system $\Sigma$ ($x_2(t,k_1,k_2)$ is not included due to the lack of space).  

\textbf{Example 2: } In this example, we consider a system $\Sigma$ with $L=2$ interconnected in a mixed way. Specifically, infinite interconnection in the first spatial dimension, and periodic interconnection in the second spatial dimension ($N_2 = 3$). The system matrices are chosen as 
\begin{equation}
\begin{cases}
\begin{aligned}
A_{TT} &= \begin{bmatrix}
-1 & 0\\ 0 & -1
\end{bmatrix},~A_{TS} = \begin{bmatrix}
1 & 0 & 0 & 0\\ 0 & 0 & -0.5 & 0
\end{bmatrix},\\
A_{ST} &=\begin{bmatrix}
0 & 0.5 \\1 & 0 \\ 0.5 & 0\\ 0 & 0
\end{bmatrix},~A_{SS}=\begin{bmatrix}
0 & 0 & 0 & 0\\0 & 0 & 0 & 0\\0 & 0 & 0 & 0\\0 & 0 & 0 & 0
\end{bmatrix}.
\end{aligned}
\end{cases}
\end{equation}

Applying the method proposed in Section \ref{Sec3_2}, we obtain
\begin{equation}
\begin{cases}
\begin{aligned}
\bar{e}_{0,0}(\mathbf{z}) &= 1,\\
\bar{e}_{1,0}(\mathbf{z}) &= 4,\\
\bar{e}_{2,0}(\mathbf{z}) &= 0.25z^{-1}_1z^{-1}_2+20+0.25z_1z_2,\\
\bar{e}_{3,0}(\mathbf{z})&=r_1(\mathbf{z})+ 63.875+\overline{r_1(\mathbf{z})},\\
\bar{e}_{4,0}(\mathbf{z})&=r_2(\mathbf{z})+263.4922+\overline{r_2(\mathbf{z})},
\end{aligned}
\end{cases}
\end{equation}
where 
\begin{equation*}
\begin{aligned}
r_1(\mathbf{z}) &=  0.0625z^{-2}_1z^{-2}_2+4z^{-1}_1z^{-1}_2,\\
r_2(\mathbf{z}) &=  0.03125z^{-3}_1z^{-3}_2+2.2539z^{-2}_1z^{-2}_2+48.2188z^{-1}_1z^{-1}_2,
\end{aligned}
\end{equation*}
and some rounding approximations are employed due to the lack of space. It is found that $\mathcal{N}=\emptyset$, thus the first sub-condition in Theorem \ref{thm2} holds. 

\begin{figure}
	\begin{center}
		\includegraphics[height=6cm]{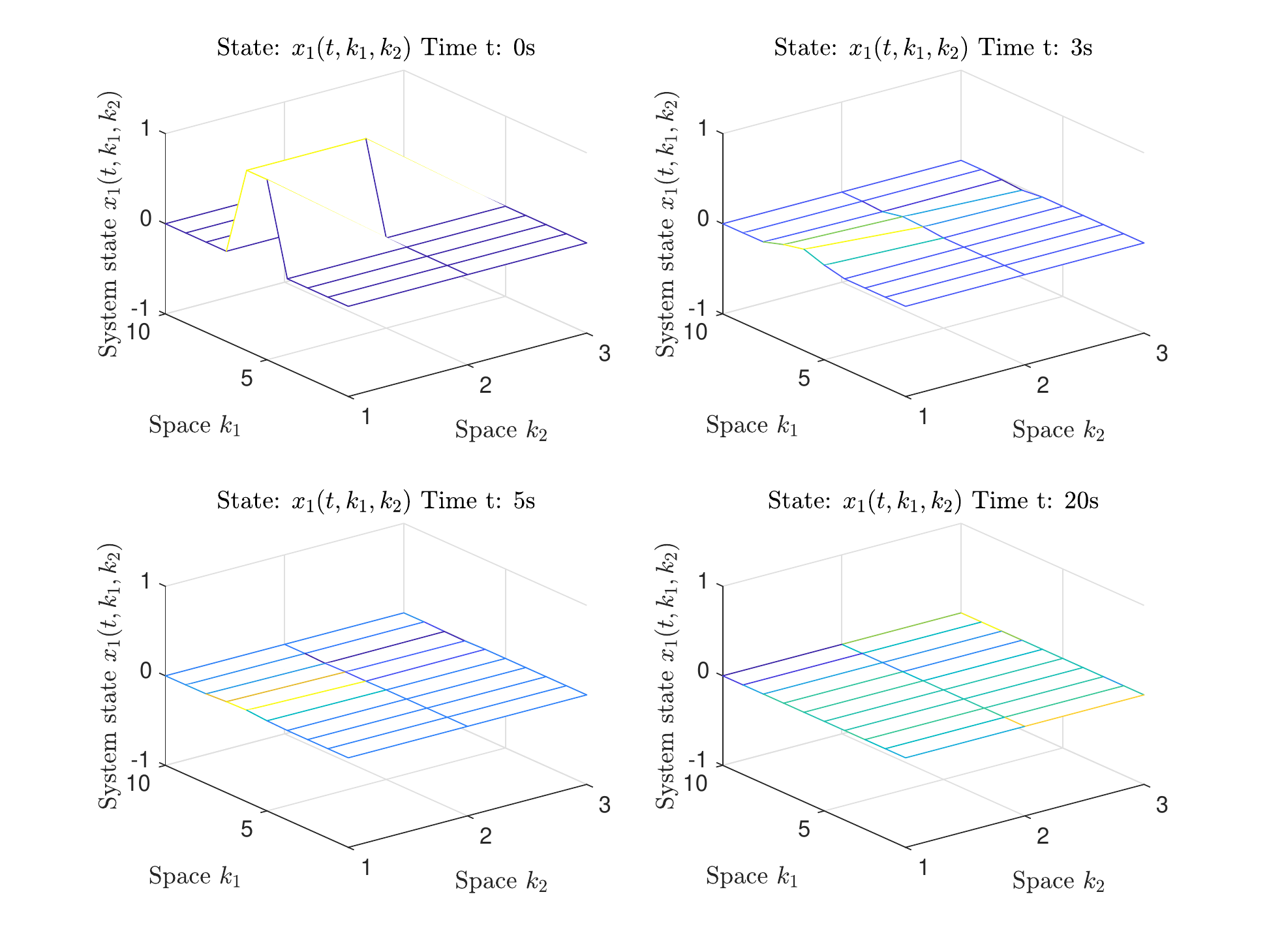}    
		\caption{State response $x_{1}(t,k_1,k_2)$ at different time}  
		\label{fig:ex3x1}                                 
	\end{center}                                 
\end{figure}

For second sub-condition, let us define 
\begin{equation}
D_1(\mathbf{z}) = z^3_2 - 1, \forall z_2 \in \mathbb{T}.
\end{equation}
Solving SDP (\ref{opt2}) for $\bm{e}=(0,0)$, the following index $\epsilon$ is achieved
\begin{equation}
\epsilon = 19.5> 0,
\end{equation}
which implies the system $\Sigma$ considered in this example is exponentially stable. Similarly, consider the following initial condition
\begin{equation}\label{initC3}
\begin{aligned}
\forall i=1,2,~x_i(0,k_1,k_2) = \begin{cases}
1,~&\text{if } k_1=5,~k_2=1; \\
&\quad k_1=6,~k_2=1; \\
&\quad k_1=6,~k_2=2,\\
0,~&\text{otherwise,}
\end{cases}
\end{aligned}
\end{equation}
the state responses $x_{1}(t,k_1,k_2)$ at $t=0$, $3$, $5$, $20$s are presented in Fig.~\ref{fig:ex3x1}.

\section{Conclusion}\label{sec5}

In this paper, positive trigonometric polynomial techniques have been exploited to establish the stability of SISs. Three types of topological interconnections were considered for each spatial direction, including infinite interconnection, periodic interconnected structure, and spatially reversible finite extent. The stability of SISs with infinite interconnections for all spatial dimensions was first explored, and inspired by the idea of rational parameterization, the addressed problem was converted into a stability test of a constant matrix and the global positivity of a trigonometric polynomial. Then, SISs of mixed infinite and periodic interconnections were considered to encompass more general topological structures. The idea of robust stabilizability function was referred to find a trigonometric polynomial whose local positivity corresponds to the stability of interest. Both the global and local positivity of the obtained polynomials can be checked via SOS decomposition. Benefiting from some existing results on spatially reversible interconnected systems, the derived methods could be applicable to all possible topological structures constructed by the three types of interconnections considered. Finally, the generalized trace parameterization of polynomials was explored, based on which, two SDPs were proposed to quantify the derived theoretical results. Although some relaxations were employed due to computational complexity, it was seen from the numerical examples that the effect seems to be negligible.

\section*{Acknowledgments}
We thank just about everybody.

\bibliography{library}

\bibliographystyle{abbrv}

\end{document}